\documentclass[pra,preprint,showpacs]{revtex4}

\usepackage{hyperref}

\usepackage[centertags]{amsmath}
\usepackage[koi8-r]{inputenc}
\usepackage[russian]{babel}
\usepackage{amsfonts}
\usepackage{amssymb}
\usepackage{amsthm} 
\usepackage{newlfont}
\usepackage{epsfig}
\usepackage{amscd}
\usepackage{graphicx}
\usepackage{epsfig}
\usepackage{footnote}
\usepackage{lipsum}
\usepackage{color}
\usepackage{xcolor}
\usepackage{multirow}
\usepackage{subcaption}

\newcommand{\beq}{\begin{equation}}
\newcommand{\eeq}{\end{equation}}
\newcommand{\ba}{\begin{array}}
\newcommand{\ea}{\end{array}}
\newcommand{\bea}{\begin{eqnarray}}
\newcommand{\eea}{\end{eqnarray}}

\begin{document}

\title{Restoring of quantum state transferred along XY-spin chain and entanglement evolution}

\author{N.A.Tashkeev}

\address{Lomonosov Moscow State University, Moscow, 119991, Russia}

\author{A.I.Zenchuk}

\address{Federal Research Center of Problems of Chemical Physics and Medicinal Chemistry RAS,
Chernogolovka, Moscow reg., 142432, Russia}.


\begin{abstract}
We propose a protocol restoring the state transferred along the spin chain governed by the XY-Hamiltonian. Such dynamics does not preserve the excitation number and leads to mixing multiple-quantum coherence matrices of orders having the same  parities. Depending on the initial state, this results in
 generating ether all possible  or only even-order multiple-quantum   coherence matrices. Restoring is established via the special unitary transformation applied to the receiver side of the chain (extended receiver). An example of restoring  $\pm1$ and $\pm 2$-order coherence matrices in  two-qubit state transfer is considered. Entanglement transfer in such process  is also studied and possibility of its amplification is demonstrated.
\end{abstract}


\maketitle

\section{Introduction}
\label{Section:Introduction}

The problem of reliable transfer of a quantum state  was formulated in the famous paper by Bose \cite{Bose} for a homogeneous chain. However, evolution of a sender's initial  state along the homogeneous chain is not effective and associated fidelity of state transfer from  the sender to the receiver is low. There are two basic methods allowing to increase  the state-transfer fidelity, both appeared soon after the above cited paper.  First method provides  the  perfect state transfer in the completely inhomogeneous $XX$-chain with nearest neighbor interactions  \cite{CDEL,KS}. The  second method provides  the high-fidelity state transfer along the homogeneous spin chain  with remoted end-nodes \cite{GKMT}. Unlike the perfect state transfer,  the high-fidelity state transfer admits different Hamiltonians governing the spin dynamics. Moreover, the spin chain with properly adjusted end-bond  is a  model providing  stability  of the  state transfer under small perturbations of the Hamiltonian  
\cite{CRMF,ZASO,ZASO2,ZASO3}. From the practical point of view, most profitable are   photon systems providing the long distance communication 
\cite{PBGWK,PBGWK2,DLMRKBPVZBW}. Although  spin chains  are also of interest \cite{PSB,LH} and can be used in local communications between blocks of a particular device.   

Moreover, 
{the high-fidelity state transfer was also organized through the one-dimensional  ring of qubits with nearest-neighbor interactions and properly fixed coupling constants ~\cite{Osborne2004}. 
In addition, it was shown that  an increase in both speed and fidelity of state transfer can be reached by applying the  specially designed control signal to a single qubit  ~\cite{SchirmerPRA2009} or  to both  end-qubits of the one-dimensional lattice~\cite{Ashhab2015}. Notice that the control-signal method allows approaching the quantum speed limit in information transfer trough a quantum channel  \cite{OsendaPLA2021}. 

Quantum state transfer was also combined  with remote implementation of quantum gates \cite{Ashhab2012,Z_2020}, in particular, of inner product of  two vectors \cite{SZ_2019}.
We also mention the information transfer through the spin chain subjected to  irregular kicks~\cite{AubourgJPB2016}, conditional controlled state transfer using periodic driving potential in Heisenberg  spin chains ~\cite{ShanSciRep2018}, 
 quantum  control
 via cutting and stitching of exchange coupling between the spins
 ~\cite{PyshkinNJP2018}, 
 state propagation along the chain with various number of optimized  coupling constants ~\cite{FZ_2009,FerronPS2022}.  
 All the above shows that protocols  of   state transfer involve (and sometimes are based on) elements of remote state control \cite{Koch2022,KuprovBook} which 
 attracts attention of many researchers.

Our paper  is devoted to the quantum state restoring  \cite{FZ_2017,BFZ_2018,Z_2018} resulting to optimized (in certain sense) state transfer \cite{FPZ_2021,BFLPZ_2022,FPZ_2024,BDFKLPZ_2024,TZ_2024} which is one of approaches to the above problem of state (information) transfer applicable mainly to the mixed states. 
 By optimized state transfer we mean the protocol resulting to the  elements of the receiver's density matrix $r$  proportional to the appropriate elements of the sender's initial density matrix $s$ \cite{FPZ_2021} (state restoring):
\begin{eqnarray}\label{restore}
r_{ij} = \lambda_{ij} s_{ij}, 
\end{eqnarray}
up to the diagonal elements which can not be given the  form (\ref{restore}) because of the trace-normalization condition  imposed on the density matrix. In addition, the coefficients $\lambda_{ij}$ ($\lambda$-parameters) are optimized by maximizing the minimal absolute value of them, i.e. by maximizing the parameter $\Lambda$,
\begin{eqnarray}\label{Lambda}
\Lambda=\min_{i,j} |\lambda_{ij}|.
\end{eqnarray}
Such optimization as well as the state restoring (\ref{restore}) can be achieved due to the specially  selected parameters of the unitary transformation applied
to  the so-called extended receiver  represented by the nodes of the receiver together with several neighboring nodes.

{We emphasize that the state restoring doesn't require large values for the $\lambda$-parameters to completely recover the transferred state, which is an advantage of state restoring in comparison with high-fidelity state transfer where the fidelity between the states $s$ and $r$ must be close to 1 in order to avoid the significant deformation of the transferred state. To obtain the nondiagonal elements $s_{ij}$ after state restoring, one has to measure $r_{ij}$ via the quantum state tomography and then divide the result by $\lambda_{ij}$. From this standpoint, the state-restoring is a hybrid algorithm allowing the perfect state recovering via the mixed quantum-classical tool. This is possible due to the universality of the $\lambda$-parameters that are the same for transferring any state $s$ and therefore are included into the state-restoring protocol. The problem of restoring the diagonal elements can be partially overcame by selecting such diagonal elements (or, more generally, the elements of the 0-order coherence matrix) that can be perfectly transferred (up to swapping two diagonal elements), as shown in \cite{BFLPZ_2022}. Thus, the state restoring is split into two subtasks: (i) restoring the non-zero order coherence matrices and (ii) constructing the 0-order coherence matrix that can be perfectly transferred. Or, another way: (i) restoring all non-diagonal elements of the whole density matrix and (ii) constructing the diagonal elements that can be perfectly transferred. Since the problem (ii) is essentially the same as in the case of evolution preserving the excitation number considered earlier, we do not refer to this problem in our paper and concentrate on describing the new features of the state-restoring protocol.}

In the recent papers \cite{FPZ_2021,BFLPZ_2022,FPZ_2024,BDFKLPZ_2024,TZ_2024} 
this task was simplified by using the evolution under the Hamiltonian preserving the number of excited spins \cite{FZ_2017}.  In that case, the $N$-qubit system evolves in the $N+1$-dimensions vector space rather than in $2^N$-dimensional space in generic case.  Moreover,  such evolution doesn't mix the elements of multiple-quantum (MQ) coherence matrices of different orders.
We recall that the  $n$-order MQ coherence matrix $\rho^{(n)}$ collects those elements of the density matrix that provide transitions in the state space with changing the $z$-projection of the total spin-momentum    by $n$. Direction $z$, for instance, can be the direction of the external magnetic field  if it is involved in the process.  Therefore, the $N$-qubit  density matrix $\rho$ can be represented as the sum of  MQ coherence matrices:
\begin{eqnarray}\label{mq}
\rho=\sum_{n=-N}^N \rho^{(n)}.
\end{eqnarray}

In our paper we consider another case of state evolution governed by the $XY$-Hamiltonian that doesn't preserve the excitation number in the system. In this case evolution mixes the MQ-coherence matrices of the orders having the same parity, {i.e, this Hamiltonian still supports certain symmetry in the system}. Such mixing complicates restoring protocol increasing the number of restoring conditions. Since restoring can not be applied to all elements of the density matrix in general (at least, the diagonal elements can not be restored) and $\lambda$-parameters are less than unit, we introduce proper characteristics to evaluate the quality of considered protocol. {Partially, the choice of the $XY$-Hamiltonian is due to the fact that this Hamiltonian can be created in the multiple-quantum (MQ) NMR experiment using the special series of magnetic pulses.  Therefore, we propose a protocol for state restoring in $XY$-spin evolution and describe its features associated with mixing multiple-quantum coherence matrices with the same order-parity keeping in mind (for further study) that this Hamiltonian can be naturally treated as a particular case of the completely anisotropic $XY$-Hamiltonian.}

For clarity, we consider the two-qubit  state transfer along the six-qubit chain. We perform restoring  together with optimization of $\lambda$-parameters via the unitary transformation applied to the four-qubit extended receiver. For simplicity, we consider pure initial states of the sender.
As an example of quantity transferred from the sender to the receiver, we consider the entanglement between two qubits of the sender and appropriate entanglement between two qubits of the restored receiver.  The set of parameters of the pure sender's initial state can be split   into two  families of  parameters revealing, respectively,     ''strong''   and ''weak''  effect on entanglement.  As a measure of entanglement we use concurrence introduces in  Wootters criterion \cite{HW,Wootters}.

The paper is structured as follows. Sec.\ref{Section:SpinDynamics} is devoted to describing the spin dynamics along the spin chain governed by the $XY$-Hamiltonian.  We also describe the restoring protocol, fidelity-based method for fixing the time instant for state registration at the receiver site, introduce characteristics of restoring protocol and give come remarks on  evolution of even-order MQ-coherence matrices. Examples of the 2-qubit state restoring in the 6-qubit spin chain are given in Sec.\ref{Section:Example}. We also consider the transfer of 2-qubit entanglement measured by concurrence (Wootters criterion). { Discussion on unitary  restoring  transformation, which is a key object in the restoring process, is presented in Sec.\ref{Section:discussion}.}
 Basic conclusions are given in Sec.\ref{Section:Conclusions}. An important expressions for the $\lambda$-parameters in terms of  the elements of the evolution operator  are derived in Appendix, Sec.\ref{Section:Appendix}.

\section{Spin dynamics under XY-Hamiltonian}
\label{Section:SpinDynamics}
We consider the spin dynamics 
 governed by the homogeneous XY-Hamiltonian with the nearest-neighbor interactions:
\begin{eqnarray}\label{H}
H&=& \sum_{i=1}^{N-1}{D(I_{i,x}I_{i+1,x}-I_{i,y}I_{i+1,y})},
\end{eqnarray}
where $D$ 
 is the coupling constant between the nearest nodes,
$ I_{i\alpha}(\alpha=x, y, z)$ is the projection operator of the $i$th spin on the $\alpha$-axis.
Unlike the XX-Hamiltonian considered in our previous papers on quantum state restoring \cite{FZ_2017,Z_2018,FPZ_2021}, this Hamiltonian doesn't commute with the z-projection operator of the total spin momentum,
\begin{eqnarray}
\label{com}
&& [H,I_{z}]\neq 0, \;\;\;I_{z}=\sum_{i}{I_{iz}},
\end{eqnarray}
and therefore does not preserve the excitation number during the evolution. Nevertheless,  this Hamiltonian still possesses certain symmetry.
In the basis 
\begin{eqnarray}\nonumber
&&
B^{(0)}=\{|0\rangle\}, \; B^{(1)}=\{ |n_1\rangle\}, \;B^{(2)}=\{|n_1n_2\rangle\}, \;\dots, \\\nonumber
&&
B^{(N-1)}=\{ |n_1\dots n_{N-1}\rangle\}, \; B^{(N)}=\{ |\underbrace{1\dots 1}_N\rangle\},  \\\label{basis}
&&0<n_1<n_1\dots<n_{N-1}\le N
\end{eqnarray}
(here $B^{(n)}$ denotes the set of $n$-excitation basis vectors and $n_i$, $i=1,\dots, N-1$, is the position of the excited spin)
 the Hamiltonian has the following block-structure:
\begin{eqnarray}\label{HB}
H=\left(
\begin{array}{cccccccc}
0^{(00)}& 0^{(01)} & H^{(02)}& 0^{(03)} & \cdots&0^{(0,N-2)}&0^{(0,N-1)}& 0^{(0N)}\cr
0^{(10)}&0^{(11)}& 0^{(12)} & H^{(13)}& \cdots & 0^{(1,N-2)}&0^{(1,N-1)}&0^{(1N)}\cr
H^{(20)}&0^{(21)}& 0^{(22)} & 0^{(23)}& \cdots & 0^{(2,N-2)}&0^{(2,N-1)}&0^{(2N)}\cr
\cdots&\cdots&\cdots&\cdots&\cdots &\cdots&\cdots &\cdots\cr
0^{(N0)}& 0^{(N1)}  & 0^{(N2)}& 0^{(N3)} & \cdots&H^{(N,N-2)}&0^{(N,N-1)}&0^{(NN)} 
\end{array}
\right),
\end{eqnarray}
here only blocks $H^{(nm)}= H^{(mn)}$, $|n-m| = 2$, are nonzero.
Each $(nm)$-block in the Hamiltonian transfers   $n$-excitation states to   $m$-excitation ones. Its dimensionality is $C^n_N \times  C^m_N$.
Then, the evolution operator $V=e^{-iHt}$ has the block structure (for even $N$)
\begin{eqnarray}\label{VB}
V=\left(
\begin{array}{cccccc}
V^{(00)}& 0^{(01)} & V^{(02)}& 0^{(03)} & \cdots&V^{(0N)}\cr
0^{(10)}&V^{(11)}& 0^{(12)} & V^{(13)}& \cdots & 0^{(1N)}\cr
V^{(20)}&0^{(21)}& V^{(22)} & 0^{(23)}& \cdots & 0^{(2N)}\cr
\cdots&\cdots&\cdots&\cdots&\cdots &\cdots\cr
V^{(N0)}& 0^{(N1)}  & V^{(N2)}& 0^{(N3)} & \cdots&V^{(NN)} 
\end{array}
\right)
\end{eqnarray}
In (\ref{VB}), only the blocks $V^{(nm)}$, $n-m=0 \mod 2$ are nonzero.  
Such evolution  leads to mixture of only the MQ-coherence matrices of the same parity. Below we use the dimensionless time $\tau = D t$. 

Similar to Refs.\cite{FZ_2017,Z_2018,FPZ_2021} on restoring  states transferred along the spin chain governed by $XX$-Hamiltonian, we consider 
 the communication line consisting of  the $N^{(S)}$-node sender $S$ (where the needed state is installed),  the $N^{(R)}$-node receiver $R$ (where the transferred state is restored)  and the $N^{(TL)}$-node transmission line  $TL$ connecting the sender to the receiver. In addition, the receiver is 
 embedded into the  extended receiver $ER$ to which the unitary restoring transformation is applied.

\subsection{State restoring}

At some time instant $\tau$, we apply a unitary transformation $U(\varphi)$ (which involves the set of free parameters $\varphi = \{\varphi_1,\varphi_2,\dots\}$) to the extended receiver. This transformation must have the same block-structure  (\ref{VB}) to keep the evolution symmetry (even $N^{(ER)}$):
\begin{eqnarray}\label{ERB}
U=\left(
\begin{array}{cccccc}
U^{(00)}& 0^{(01)} & U^{(02)}& 0^{(03)} & \cdots&U^{(0N^{(ER)})}\cr
0^{(10)}&U^{(11)}& 0^{(12)} & U^{(13)}& \cdots & 0^{(1N^{(ER)})}\cr
\cdots&\cdots&\cdots&\cdots&\cdots &\cdots\cr
U^{(N^{(ER)}0)}& 0^{(N^{(ER)}1)}  & U^{(N^{(ER)}2)}& 0^{(N^{(ER)}3)} & \cdots&U^{(N^{(ER)}N^{(ER)})} 
\end{array}
\right),
\end{eqnarray}
 Thus, the total unitary transformation $W$ of the initial state,
 \begin{eqnarray}\label{W}
W(\tau)&=&\Big(I_{S,\overline{TL}}\otimes U\Big) V(\tau)
\end{eqnarray}
 has also a block-diagonal  form (\ref{VB}) up to the replacement $V\to W$.
Here $I_{S,\overline{TL}}$ is the identity matrix in the space $S\cup \overline{TL}$,  $\overline{TL}$ means the $TL$ without nodes of the $ER$ (thus, the whole system is $S\cup TL \cup R \equiv S\cup\overline{TL}\cup ER$).
Therefore, 
\begin{eqnarray}\label{rhot0}
\rho(\tau)= W(\tau) \rho(0) W^\dagger (\tau),
\end{eqnarray}
or, for  the blocks, 
\begin{eqnarray}\label{rhot1}
&&
\rho^{(nm)} (\tau)= \sum_{l-k=p\!\!\! \mod 2}W^{(nk)}(\tau) \rho^{(kl)}(0) (W^\dagger)^{(lm)} (\tau),\\\label{nmjk}
&&m-n=p \!\!\!\mod 2,\;\;k-n =0  \!\!\!\mod 2,\;\;
 m-l=0  \!\!\!\mod 2,\;\; p=0,1.
\end{eqnarray}
We consider the following tensor-product initial state:
\begin{eqnarray}\label{inst}
\rho(0)=s\otimes \rho^{(TL,R)}(0),
\end{eqnarray}
where $s$ is an arbitrary $N^{(S)}$-qubit sender's initial state and $ \rho^{(TL,R)}(0)$ is the initial state of the joined subsystem $TL\cup R$ which is a $0$-excitation state:
\begin{eqnarray}
 \rho^{(TL,R)}(0) =|0\rangle_{R,TL} \; _{R,TL}\langle 0|.
\end{eqnarray}

We represent $U$ in terms of  exponents \cite{Z_2018},
\begin{eqnarray}\label{exponent}
U(\varphi)=\prod_{j} e^{iA_j(\varphi_j)},\;\; A_j^\dagger =A_j,
\end{eqnarray}
where each $A_j$ is the $2^{N^{(ER)}}\times 2^{N^{(ER)}}$ Hermitian off-diagonal matrix which  involves only two nonzero elements introducing  the real parameter
$\varphi_j$. There are 
\begin{eqnarray}\label{varphi}
P=\sum_{n-m=0\!\!\! \mod 2}  C^n_{N^{(ER)}}  C^m_{N^{(ER)}}- 2^{N^{(ER)}}
\end{eqnarray}
(where we subtract the number of diagonal elements of $U$)
linearly independent matrices of this type, therefore there are $P$  real parameters $\varphi_i$ in the unitary transformation $U$, 
\begin{eqnarray}\label{Udim}
\varphi=\{\varphi_i:\;i=1,\dots,  P\}.
\end{eqnarray}
This setup is similar to that given in
\cite{Z_2018} and used latter in \cite{FPZ_2021,BFLPZ_2022,TZ_2024}.

The density matrix  of the receiver at some time instant $\tau$  is obtained from $\rho$ by partial tracing over $S$ and $TL$:
\begin{eqnarray}\label{rhot}
r(\tau)=Tr_{S,TL} \big(\rho(\tau)\big) = Tr_{S,TL} \big(W(\tau)\rho(0) W^{\dagger}(\tau)\big),
\end{eqnarray}
or, for blocks,
\begin{eqnarray}\label{rhot3}
&&
r^{(nm)}(\tau)= Tr_{S,TL}\sum_{l-k=p \!\!\! \mod 2} \big(W^{(nk)}(\tau)\rho^{(kl)}(0) (W^{\dagger})^{(lm)}(\tau)\big),
\end{eqnarray}
where superscripts $n,\;m,\;k,\;l$ satisfy (\ref{nmjk}).
Both initial sender's state and the receiver's state have similar structures:
\begin{eqnarray}\label{s}
 s&=& \left(
\begin{array}{cccc}
s^{(00)}& s^{(01)}&\cdots& s^{(0N^{(S)})} \cr
s^{(10)}&s^{(11)} &\cdots&s^{(1N^{(S)})}\cr
\cdots & \cdots & \cdots & \cdots \cr
s^{(N^{(S)}0)}&s^{(N^{(S)}1)} &\cdots&s^{(N^{(S)}N^{(S)})}\end{array}
\right),\;\; \\\label{r}
r&=& \left(
\begin{array}{cccc}
r^{(00)}& r^{(01)}&\cdots& r^{(0N^{(R)})} \cr
r^{(10)}&r^{(11)} &\cdots&r^{(1N^{(R)})}\cr
\cdots & \cdots & \cdots & \cdots \cr
r^{(N^{(R)}0)}&r^{(N^{(R)}1)} &\cdots&r^{(N^{(R)}N^{(R)})}\end{array}
\right)
\end{eqnarray}
and we set $N^{(S)}=N^{(R)}$ hereafter. 
Here $s^{(nm)}$ and $r^{(nm)}$ are the $C^n_{N^{(S)}}\times  C^m_{N^{(S)}}$  blocks responsible for transitions from 
$n$-excitation states to  $m$-excitation ones. 
In view of initial state (\ref{inst}), eq.(\ref{rhot3}) yields
\begin{eqnarray}\label{rr}
r^{(nm)}_{ij} &=& \sum_{ l- k = p\!\!\!  \mod 2}
\sum_{\tilde i = 1}^{C^k_{N^{(S)}}} \sum_{\tilde j = 1}^{C^l_{N^{(S)}}} 
  \lambda^{(nmkl)}_{ij \tilde i \tilde j} s^{(kl)}_{\tilde i \tilde j},
\end{eqnarray}
where superscripts $n,\;m,\;k,\;l$ satisfy (\ref{nmjk}) and we remove $\tau$ from the arguments.
Note that, unlike the case of evolution conserving the excitation number (where the single element of the highest order MQ coherence matrix has a restored form before applying the  unitary restoring  transformation to the extended receiver), now the highest-order  coherence matrix (i.e., $\pm N^{(S)}$-order) is to be restored as well because of mixing of all 
MQ-coherence matrices with the same order-parity.

We require
\begin{eqnarray}\label{lam}
 \lambda^{(nmkl)}_{ij \tilde i \tilde j}=\lambda^{(nm)}_{ij} \delta_{i\tilde i} \delta_{j\tilde j}\delta_{nk}\delta_{ml},
\end{eqnarray}
which is equivalent to the following system
\begin{eqnarray}\label{lameq}
&&
 \lambda^{(nmkl)}_{ij \tilde i \tilde j}=0, \;\;  \{nmij\} \neq  \{kl\tilde i  \tilde j\},
\end{eqnarray}
with relations between indexes (\ref{nmjk}) and $l-k=p\!\!\!\mod 2$.
We call system (\ref{lameq}) the system of restoring conditions.

System (\ref{lameq}) can be split into two families of equations corresponding to $p=0$  ($m-n$ is even) and $p=1$  ($m-n$ is odd) in (\ref{nmjk}). 
Blocks with odd $m-n$ can be completely restored. There are 
\begin{eqnarray}
A^{(odd)} = \sum_{n-m =1  \!\!\!\mod 2} C^{(n)}_{N^{(R)}}  C^{(m)}_{N^{(R)}} 
\end{eqnarray}
elements in these blocks. Each element is a linear combination of $A^{(odd)}$ elements of $s$.  Therefore, to  keep the structure  (\ref{lam}), we have to remove 
$A^{(odd)}-1$ elements of $s$ from each of $A^{(odd)}$ elements of $r$.
For that, we have to solve $A^{(odd)}(A^{(odd)}-1)$ equations of type (\ref{lameq})  for  the parameters $\varphi$ of  the unitary transformation of  the extended receiver. 

Regarding the blocks with even $m-n$, we have to select blocks of zero-order  coherence matrix, there are
\begin{eqnarray}
A^{(0)} = \sum_{n=0}^{N^{(R)}} (C^n_{N^{(R)}})^2
\end{eqnarray}
elements in those blocks.
All other blocks with even $m-n$  contain
\begin{eqnarray}
A^{(even)} = \sum_{{n-m =0  \!\!\!\mod 2}\atop{n\neq m}} C^{(n)}_{N^{(R)}}  C^{(m)}_{N^{(R)}} 
\end{eqnarray}
elements.  Each of these element is a linear  combination of $A^{(even)}+A^{(0)}$ elements of $s$.
Thus, we have to remove   $(A^{(even)}+A^{(0)}-1)$  extra terms from $A^{(even)}$ elements. 
To this end  we have to solve $A^{(even)} (A^{(even)}+A^{(0)}-1)  $   equations of type (\ref{lameq})  for  the parameters $\varphi$ of  the unitary transformation 
applied  to the extended receiver at the selected time instant $\tau_0$ (see Sec.\ref{Section:timeinst}).  We emphasize that the unitary transformation of the extended receiver constructed in this way doesn't depend on the particular sender's state to be transferred and restored and therefore  it is universal, similar to  the unitary transformations restoring the transferred state along the $XX$-Hamiltonian  \cite{FPZ_2021}.

All in all, the elements of the  restored state $r$ are related to the appropriate elements of the initial state $s$ at the   time instant $\tau_0$ (selected for state registration) as follows:
\begin{eqnarray}\label{restored}
r^{(nm)}_{ij}(\tau_0) = \lambda^{(nm)}_{ij}(\tau_0) s^{(nm)}_{ij}, \;\;m-n=\pm 1,\pm 2, \dots.
\end{eqnarray}
We emphasize that the restoring   elements of zero order coherence matrix (blocks $r^{(nn)}(\tau_0)$, $n=0,\dots N^{(R)}$)  is not considered here, although the nondiagonal elements of this matrix might be restored in a similar way.

\subsection{Time instant for state registration}
\label{Section:timeinst}

Recall  that  not all elements of the sender's density matrix can be restored in the receiver's state. Let the set ${\cal{R}}$ collect the $\lambda$-parameters of the restored elements,
i.e. 
\begin{eqnarray}\label{R}
&&{\cal{R}}(\varphi,\tau)=\{\lambda_{ij}: r_{ij} = \lambda_{ij}(\varphi,\tau) s_{ij}\},\\\label{Lambda2}
&&
\Lambda=\min_{|\lambda_{ij}|} {\cal{R}},
\end{eqnarray}
 where $\varphi$ (given in (\ref{Udim})) is the list of  parameters of the unitary transformation applied to the extended receiver, and parameter $\Lambda$ was mentioned in (\ref{Lambda}). 
We call the set $(\varphi,\tau)$ optimal, $(\varphi_{opt},\tau_{opt})$, if $\Lambda$ is maximized at $\varphi_{opt}$, $\tau_{opt}$, i.e.
\begin{eqnarray}
\label{opt}
\Lambda(\varphi_{opt},\tau_{opt}) = \max_{\tau,\varphi} \min_{|\lambda_{ij}|} {\cal{R}}(\varphi,\tau).
\end{eqnarray}
Here we consider such parameters $\varphi$ that solve the restoring conditions (\ref{lameq}). Thus, time $\tau$ must, in general, be included into the set of controlling parameters along with $\varphi$,  {like it  is done in Refs.\cite{BFLPZ_2022,FPZ_2024}. But the time instant in the list of optimizing parameters make calculations more cumbersome. However, it was shown in Ref.\cite{DZ_2017} that the time instant maximizing the probability of the excited-state transfer coincides with the time instants maximizing the values of certain other characteristics of state/information transfer used in that paper. This motivates us to} select the optimal time instant for state registration as the time instant that maximizes the fidelity of a pure $N^{(S)}$-qubit state transfer  ($s=|\psi\rangle \langle \psi|$) averaged over the sender's initial states $|\psi\rangle$ before applying the  unitary restoring transformation $U$, i.e. $\varphi=0$:
\begin{eqnarray}\label{F}
F(\tau)=
\Big\langle \langle \psi|r(\tau)|_{\varphi=0}|\psi\rangle \Big\rangle_{|\psi\rangle},
\end{eqnarray}
In other words, 
\begin{eqnarray}\label{t0}
F(\tau_0) = \max_{0\le\tau \le T} F(\tau),
\end{eqnarray}
where $T$ is a large enough time interval. Therefore, we replace definition (\ref{opt}) of the optimized parameters with the following one performed at the time instant $\tau_0$ fixed in Eq.(\ref{t0}):
\begin{eqnarray}
\label{opt2}
\Lambda(\varphi_{opt},\tau_0) = \max_{\varphi} \min_{|\lambda_{ij}|} {\cal{R}}(\varphi,\tau_0).
\end{eqnarray}
{In any case, optimization of time-instant for state registration is a standard requirement and  a particular method chosen for fixing this time instant  can yield only some quantitative correction to the obtained result of state restoring.} Hereafter we write $\tau$ instead of $\tau_0$ for simplicity.
\subsection{Characteristics of  state restoring}

According to the above discussion, the restored state differs from the initial sender's state although repeats certain features of its structure. 
Therefore, fidelity doesn't properly describe the effectivity of restoring because it has no information about the structure of the restored state. 
Therefore, we need to introduce different characteristics which would reflect the following features of restoring.
\begin{itemize}
\item[-]
Not all elements are restored.
\item[-]
$\lambda$-parameters are less then unit by absolute value.
\end{itemize}
These features prompt us to  introduce the following parameters.

\begin{enumerate}
\item The fraction of restored elements of $n\times n$ matrix:
\begin{eqnarray}
N_r =\frac{N_{\cal{R}}}{n^2},
\end{eqnarray} 
where $N_{\cal{R}}$ is the length of ${\cal{R}}$ in (\ref{R}).
\item The minimal by absolute value $\lambda$-parameter $\Lambda$ given in (\ref{Lambda}) or (\ref{Lambda2}).
\item The mean absolute value of $\lambda$-parameters:
\begin{eqnarray}
\Lambda_{avr} = \frac{\sum_i{|\cal{R}}_i|}{N_{\cal{R}}}.
\end{eqnarray}
Here ${\cal{R}}_i$ is the $i$th element of the list ${\cal{R}}$ (\ref{R}).
\end{enumerate}
The parameter $N_r$ reaches its maximal value for a given $n\times n$ matrix $r$ if the list ${\cal{R}}$ includes all non-diagonal elements. In this case
\begin{eqnarray}
N_r=\frac{n(n-1)}{n^2},\;\;\lim_{n\to\infty} N_r=1,
\end{eqnarray}
i.e. $N_r$ asymptotically reaches the perfect value although the diagonal elements (the classical part of information) can not be restored. 

The parameter $\Lambda$ means the worst $\lambda$-parameter in structural restoring. Its small value informs that the restoring protocol  is desired to be improved.

The parameter $\Lambda_{avr}$ indicates the effectiveness of state restoring. Its meaning is similar to the meaning of $\Lambda$, but it takes into account all $\lambda$-parameters. 

It is clear that both $\Lambda$ and $\Lambda_{avr}$ tend to unit  in the perfect case.

\subsection{Evolution of system with even-order MQ-coherence matrices}
Since the evolution doesn't mix odd- and even-order MQ-coherence matrices we can remove  odd-order coherence matrices from the sender's initial state and obtain the receiver's density matrix including only even-order coherence matrices. Thus,  the structure of  density matrix becomes much simpler. In formulae (\ref{s}) and (\ref{r}) only blocks $s^{(nm)}$, $r^{(nm)}$ with $m-n=0 \mod 2$ are nonzero.  System (\ref{restored}) becomes 
\begin{eqnarray}\label{restored2}
r^{(nm)}_{ij}(t) = \lambda^{(nm)}_{ij}(t) s^{(nm)}_{ij}, \;\;m-n=\pm 2, \pm 4\dots.
\end{eqnarray}
The list of parameters to be optimized is
\begin{eqnarray}
{\cal{R}} =\{ \lambda^{(nm)}_{ij}: r^{(nm)}_{ij} = \lambda^{(nm)}_{ij} s^{(nm)}_{ij},  \;m-n=\pm 2, \pm 4\dots\}.
\end{eqnarray}
Therefore, the parameter $N_r$ for   the  restored matrices in this  section is approximately twice less then this parameter in the case of restoring of all MQ-coherence matrices (except  for the diagonal elements). 

Notice that the density matrix without even-order coherence matrices does not exist because of the trace-normalization which requires presence of, at least, the 0-order coherence matrix. Therefore, in the  case of evolution not preserving the excitation number, the evolutionary density matrix consists of the MQ-coherence matrices of either all possible orders or even orders.

 \section{Example: two-qubit state transfer and restoring, entanglement amplification}
 \label{Section:Example}
 We consider two examples of state restoring. First one is related to the sender's initial state including all MQ coherence matrices, while the second example describes restoring of the matrix containing only even-order MQ coherence matrices.  We consider the short chain of six qubits  (dealing with a $2^6\times 2^6$ density matrix) with two-qubit sender and receiver and fix time instant  according to Sec.\ref{Section:timeinst} for both examples. For the restoring and optimization protocol,  we take the four-node extended receiver, then the number of real parameters in the unitary transformation applied to this extended receiver is $P=112$ according to (\ref{varphi}).
 
 \subsection{Restoring of $\pm 1$- and $\pm 2$-order coherence matrices}
 \label{Section:example1}
 For simplicity of generating $s$ we consider only pure initial states of sender.
 In this case we take 
 \begin{eqnarray}\label{Favr}
&& |\psi\rangle =
 \left(
 \begin{array}{c}
 \sin\frac{\varphi_0}{2}  \sin\frac{\varphi_1}{2}   \sin\frac{\varphi_2}{2}   \cr
e^{i \chi_1} \cos\frac{\varphi_0}{2}  \sin\frac{\varphi_1}{2}   \sin\frac{\varphi_2}{2}  \cr
 e^{i \chi_2 } \cos\frac{\varphi_1}{2}   \sin\frac{\varphi_2}{2} \cr
 e^{i \chi_3} \cos\frac{\varphi_2}{2}
 \end{array}
 \right),\\\label{ss}&& s=|\psi\rangle \langle \psi| = \left(
\begin{array}{cccc}
s_{11}&s_{12}&s_{13}&s_{14}\cr
s_{12}^*&s_{22}&s_{23}&s_{24}\cr
s_{13}^*&s_{23}^*&s_{33}&s_{34}\cr
s_{14}^*&s_{24}^*&s_{34}^*&s_{44}
\end{array}
\right),
 \end{eqnarray}
all $s_{ij}$ are expressed in terms of the parameters $\varphi_i$, $i=0,1,2$ and $\chi_i$, $i=1,2,3$.
 Using formula (\ref{F}), we average the state-transfer fidelity over the phase parameters  $\chi_i$, $i=1,2,3$, $0\le\chi_i\le 2\pi$,  and over the 4-dimensional hypersphere with the angular-parameters $\varphi_0, \varphi_1, \varphi_2$ of the initial state (\ref{Favr}). 
Thus, formula (\ref{F}) can be given the form
 \begin{eqnarray}\nonumber
 &&
 F=\frac{1}{\Omega}   \int_{0}^{2\pi} d\chi_1 \int_{0}^{2\pi} d\chi_2 \int_{0}^{2\pi} d\chi_3 \int_{0}^{2\pi} d \varphi_0   \int_{0}^{\pi} d \varphi_1  \int_{0}^{\pi} d \varphi_2  \\
 &&\langle \psi| r(\tau)|\psi\rangle J(\varphi_1,\varphi_2),
\\\label{Omega}
&&
\Omega=
 { \int_{0}^{2\pi}  d\chi_1 \int_{0}^{2\pi} d\chi_2 \int_{0}^{2\pi} d\chi_3 \int_{0}^{2\pi} d \varphi_0   \int_{0}^{\pi} d \varphi_1  \int_{0}^{\pi} d \varphi_2 J(\varphi_1,\varphi_2)}= 16\pi^5,
 \end{eqnarray}
where the Jacobian $J$ reads
\begin{eqnarray}\label{J}
J= \sin\, \varphi_1 \sin^2\,\varphi_2.
\end{eqnarray}
 The time dependence of $F$ is given in Fig.\ref{Fig:fig1}. In this figure, the optimal time instant is $\tau_0=55.5352$.
\begin{figure}[ht]
\centering
    \includegraphics[width=0.6\textwidth]{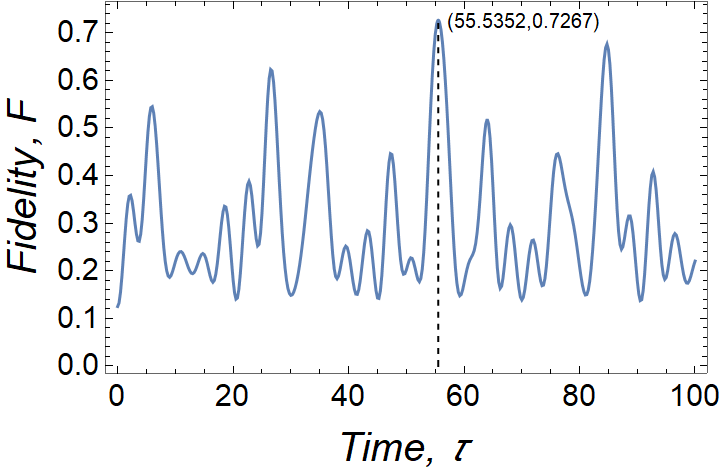}
    \caption{The time-dependence of fidelity $F$ and the preferable time instant for state registration  $\tau_0=55.5352$.}
       \label{Fig:fig1}
  \end{figure}
 
 Restoring system  (\ref{lameq}) includes 35 complex  (or 70 real) equations, which is less than the number of parameters $P=112$ in the unitary transformation. Therefore,  the solution of the restoring system is not unique. 
 We find 1000 different solutions of the restoring system  and perform optimization over these solutions with the purpose to maximize the parameter $\Lambda$ (\ref{opt2}).  
 As the result, such optimization yields the following receiver's density matrix:
 {\small \begin{eqnarray}\label{rhor1}
 &&r =\\\nonumber
 && \left(
 \begin{array}{cccc}
 r_{11} & 0.7566 e^{0.4443 i}s_{12} &0.7060 e^{-2.6881 i} s_{13}& 0.4326 e^{-2.6298 i}s_{14}\cr
 0.7566 e^{-0.4443 i}s^*_{12}&r_{22}&r_{23}&0.5524 e^{-3.0895 i}s_{24}\cr
 0.7060 e^{2.6881 i}s^*_{13}&r^*_{23}&r_{33}&0.5048e^{0.0698i}s_{34}\cr
 0.4326 e^{2.6298 i}s^*_{14}&0.5524 e^{3.0895 i}s^*_{24}&0.5048e^{-0.0698i}s^*_{34}&r_{44} 
 \end{array}
 \right)
 \end{eqnarray}}
 \begin{eqnarray}\nonumber
 r_{23}&=&0.1065 e^{2.8116i}s_{11} + 0.0036 e^{2.7464i} s_{22} + 0.0060 e^{-0.0571i}s_{33} +\\\nonumber
 && 0.1866 e^{-0.3595 i}s_{44} +0.8819 e^{3.1344 i}s_{23} + 0.0015 e^{2.0740 i}s^*_{23},\\\nonumber
 r_{11}&=&0.6810s_{11} + 0.0062  s_{22} + 0.1407s_{33} + 0.0502s_{44} +2{\mbox{Re}} (0.0054  e^{2.3881 i}s_{23}),\\\nonumber
  r_{22}&=&0.0950s_{11} + 0.9864  s_{22} + 0.0020s_{33} + 0.5160s_{44} +2{\mbox{Re}} (0.0004  e^{1.0816 i}s_{23}),\\\nonumber
  r_{33}&=&0.2187s_{11} + 0.0021  s_{22} + 0.8521s_{33} + 0.1142s_{44} +2{\mbox{Re}} (0.0050  e^{-1.7070 i}s_{23}),\\\label{rhor0}
 r_{44}&=&0.0052s_{11} + 0.0053 s_{22} + 0.0052s_{33} + 0.3196s_{44} +2{\mbox{Re}} (0.0045  e^{0.2032 i}s_{23}).
  \end{eqnarray}
  In this case $N_r=10/16=5/8$, $\Lambda=0.4326$ and $\Lambda_{avr} = 0.5905$.
Fidelity of this state is $F=0.2489$, it is less than the fidelity of non-restored state which is 0.7267, see Fig.\ref{Fig:fig1}. However, instead of high fidelity we have the structure of $r$ (\ref{rhor1}) that better reflects the structure of the sender's initial state $s$ (\ref{ss}) than the receiver's density matrix before restoring.

\subsubsection{Entanglement}
\label{Section:ent1}
Now we study how the state restoring effects some physical quantity transferred from the sender to the receiver. 
As an example of such quantity  we consider the entanglement between the qubits of the sender transferred into the entanglement between the qubits of the receiver. 
{We emphasize that the problem of maximizing the entanglement is beyond the scope of our paper, although the tool of unitary transformation can obviously be used for this purpose as well. However, with the purpose of increasing  entanglement, completely different unitary transformations of the extended receiver  must be used. We study the entanglement in the optimally  restored states without paying the special attention to its value. 
}

For the measure of entanglement we use the  Wootters criterion \cite{HW,Wootters} and calculate the concurrence via the formulae
\begin{eqnarray}\label{CR}
R=\sqrt{r (\sigma^{(y)} \otimes \sigma^{(y)})r^* (\sigma^{(y)} \otimes \sigma^{(y)})},\;\; \sigma^{(y)}=\left(\begin{array}{cc}
0&-i \cr
i& 0
\end{array}\right),
\end{eqnarray}
\begin{eqnarray}\label{C}
C=\max(0,2 \lambda_{max} - \sum_{i=1}^4\lambda_i),
\end{eqnarray}
where $\lambda_i$ are the eigenvalues of $R$, $\lambda_{max}$ is the maximal eigenvalue and $\sigma^{(y)}$ is the Pauli matrix.
We calculate the concurrence of the sender's state  $C_s$  and of the optimized receiver's state  $C_r$ for 100000 different pure initial sender's states  fixed by 100000 sets of random parameters $0\le \chi_i\le 2 \pi$, $i=1,2,3$, and $0\le \varphi_0 \le 2 \pi$, $0\le \varphi_i \le  \pi$, $i=1,2$,
and plot  $C_r$ in dependence on $C_s$ in Fig.\ref{Fig:fig2}. Generically, the state restoring decreases concurrence. However, we see the region $0\le C_s< 0.561$ where the amplification of concurrence is possible (i.e. $C_r>C_s$ for some initial states)  during the state transfer and restoring.   
\begin{figure}[ht]
\centering
    \includegraphics[width=0.6\textwidth]{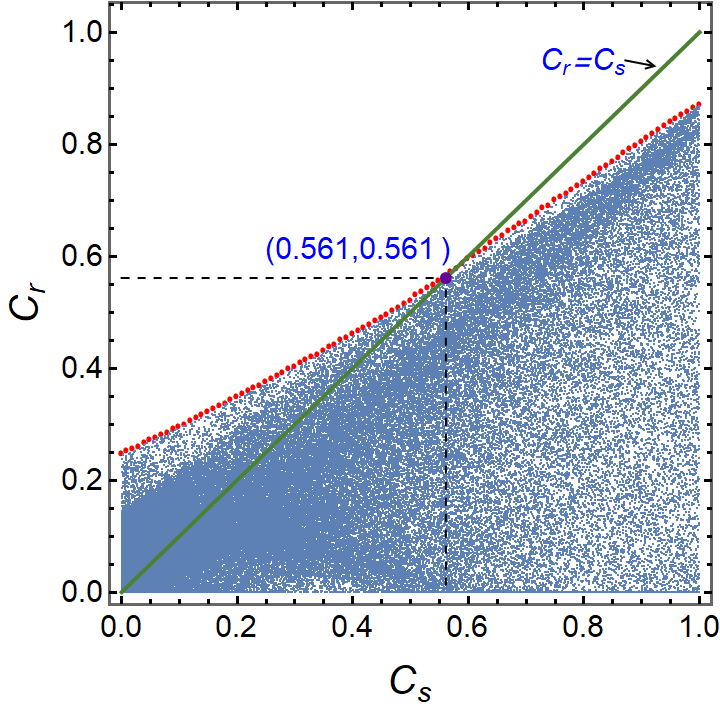}
    \caption{$N=6$; dependence of $C_r$ on $C_s$ in the two-qubit state restoring for 100000 random initial sender's states. The amplification of concurrence (i.e., $C_r>C_s$) is possible for $C_s< 0.561$. The dotted line (red online) indicates the boundary of the presented region on the plane $(C_s,C_r)$. }
    \label{Fig:fig2}
  \end{figure}
 We also shall notice that the dependence of  the concurrence on $\varphi_i$ is stronger than the dependence on $\chi_i$. To demonstrate this effect we calculate 
 six mean values of $n$th powers of concurrences  averaged over five parameters (out of six parameters $\varphi_i$ and $\chi_i$) keeping dependence on a single parameter, i.e., we introduce the quantities
 \begin{eqnarray}\nonumber
&& \langle C^n_b(\chi_i)\rangle =\frac{1}{\Omega}
   \int_{0}^{2\pi} d\chi_j \int_{0}^{2\pi} d\chi_k \int_{0}^{2\pi} d \varphi_0   \int_{0}^{\pi} d \varphi_1  \int_{0}^{\pi} d \varphi_2  C^n_b(\varphi,\chi) J(\varphi_1,\varphi_2),\\\nonumber
&&\{i,j,k\} = {\mbox{perm}} \{1,2,3\},\\\nonumber
&& \langle C^n_b(\varphi_0)\rangle =\frac{1}{\Omega}
  \int_{0}^{2\pi} d\chi_1  \int_{0}^{2\pi} d\chi_2 \int_{0}^{2\pi} d\chi_3   \int_{0}^{\pi} d \varphi_1  \int_{0}^{\pi} d \varphi_2  C^n_b(\varphi,\chi) J(\varphi_1,\varphi_2),\\\nonumber
&& \langle C^n_b(\varphi_i)\rangle=\frac{1}{\Omega}
  \int_{0}^{2\pi} d\chi_1  \int_{0}^{2\pi} d\chi_2 \int_{0}^{2\pi} d\chi_3   \int_{0}^{2\pi} d \varphi_0  \int_{0}^{\pi} d \varphi_j  C^n_b(\varphi,\chi) J(\varphi_1,\varphi_2),\\\label{CC}
&&\{i,j\} = {\mbox{perm}}\{1,2\}
 \end{eqnarray}
where $\Omega$ is defined in (\ref{Omega}), $n=1,2$, $\varphi=\{\varphi_0,\varphi_1,\varphi_2\}$, $\chi=\{\chi_1,\chi_2,\chi_3\}$ and $b$ is either $s$ or $r$. 
For $n=1$, these quantities are one-parametric mean values. For $n=2$, they are one-parametric  mean squares, which can be  used to  introduce the one-parametric mean-square deviation
\begin{eqnarray}
\delta_a C_b = \sqrt{\langle C_b^2(a)\rangle-\langle C_b (a)\rangle^2} ,
\end{eqnarray}
where $b$  is either $s$ or $r$ and $a$ is one of the parameters $\varphi_i$, $i=0,1,2$,  or $\chi_i$,  $i=1,2,3$. Functions $C_s(\varphi_i)$ and $C_r(\varphi_i)$,
\begin{eqnarray}\label{Cb}
C_b(\varphi_i) = \langle C_b (\varphi_i)\rangle \pm  \delta_{\varphi_i} C_b,\;\;i=0,\;1,\;2,
\end{eqnarray}
 are shown in Fig.\ref{Fig:Cr2}. 
The solid lines in this figure are referred to the mean values $\langle C_b (\varphi_i)\rangle$, while the error-bars denote the mean-square deviations $\delta_{\varphi_i} C_b$. We see that each mean value $\langle C_b (\varphi_i)\rangle$ significantly depend on its parameter $\varphi_i$, $i=0,\;1,\;2$. The mean-square deviations are significant and reach the value $\sim 0.3$, which is comparable with the appropriate variation interval  $\Delta_{\varphi_i} C_{b}$ 
for  $\langle C_b (\varphi_i)\rangle$, $b=s,\,r$, 
\begin{eqnarray}\label{Delta}
\Delta_{\varphi_i} C_{b}=
 \max_{\varphi_i}\langle C_{b}(\varphi_i) \rangle - \min_{\varphi_i} \langle C_{b}(\varphi_i)\rangle.
\end{eqnarray} 
\begin{figure}[ht]
\begin{subfigure}{0.45\textwidth}
   \includegraphics[width=1 \textwidth]{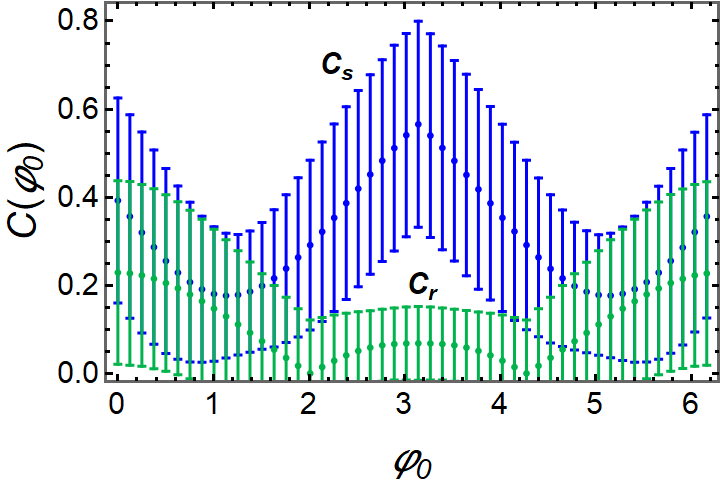}
    \caption{}
   \end{subfigure}\hspace{1cm}
    \begin{subfigure}{0.45\textwidth}
     \includegraphics[width=1\textwidth]{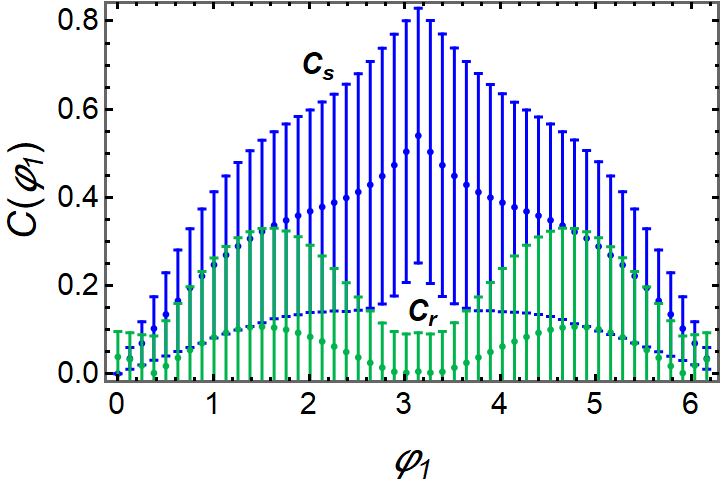}
    \caption{}
    \end{subfigure}
      \begin{subfigure}{0.45\textwidth}
     \includegraphics[width=1\textwidth]{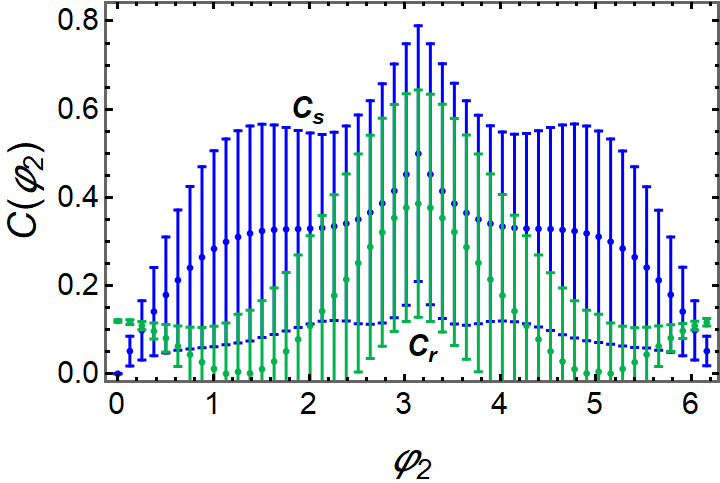}
    \caption{}
    \end{subfigure}
     \caption{ Dependence of concurrence on strong parameters $\varphi_0$, $\varphi_1$, $\varphi_2$. Here $C_s(\varphi_i)=\langle C_s(\varphi_i)\rangle  \pm \delta C_s(\varphi_i)$ (blue on line),
     $C_r(\varphi_i)=\langle C_r(\varphi_i)\rangle  \pm \delta C_r(\varphi_i)$ (green on line), $i=0,1,2$.  
(a) $C_s(\varphi_0)$,  $C_r(\varphi_0)$;
(b)  $C_s(\varphi_1)$,  $C_r(\varphi_1)$;
(c) $C_s(\varphi_2)$,  $C_r(\varphi_2)$;
The parameters $\delta_{\varphi_i}C_b$ and $\Delta_{\varphi_i}C_b$, $b=s,r$, are given in Table \ref{Table:C}.}
    \label{Fig:Cr2}
  \end{figure}

  \begin{figure}[ht]
   \includegraphics[width=0.6 \textwidth]{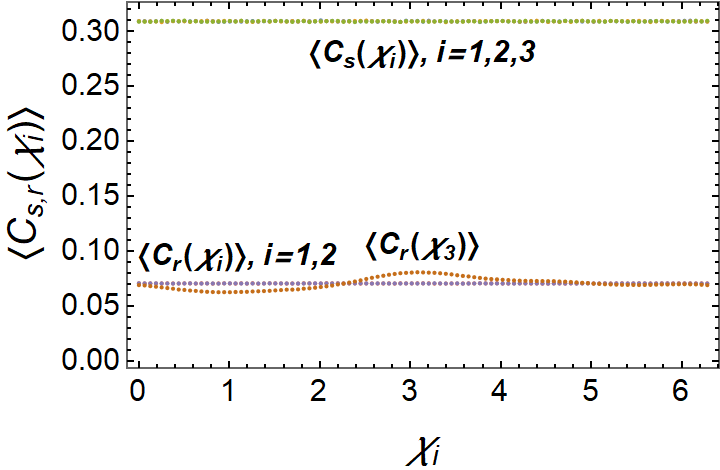}
    \caption{Dependence of concurrence on weak parameters $\chi_i$, $i=1,2,3$; functions $\langle C_s(\chi_i)\rangle$ and $\langle C_r(\chi_i)\rangle$, $i=1,2,3$, are depicted. 
    The mean-squared deviations $\delta C_s(\chi_i)$, $\delta C_r(\chi_i)$ (not shown in this figure) are almost constant for $i=1,2,3$  as represented in Table \ref{Table:C}.
}
    \label{Fig:Cr3}
  \end{figure}
  Now we turn to another set of parameters $\chi_i$, $i=1,2,3$. The mean values 
$\langle C_b(\chi_i)\rangle$ are shown in Fig.\ref{Fig:Cr3}. Unlike the mean values $\langle C_b(\varphi_i)\rangle$ shown in Fig.\ref{Fig:Cr2},  the mean values $\langle C_b(\chi_i)\rangle$, $i=1,2,3$, are constant except $\langle C_r(\chi_3)\rangle$. Moreover, 
$\langle C_s(\chi_1)\rangle =\langle C_s(\chi_2)\rangle =\langle C_s(\chi_3)\rangle   $ and $\langle C_r(\chi_1)\rangle =\langle C_r(\chi_2)\rangle$.
The mean-square deviations $\delta_bC_{\chi_i}$ (they are not shown in the graph) are also almost constant, as shown in  Table \ref{Table:C}. 
 The large mean-square deviations mean that, although the mean values of the concurrences in Fig.\ref{Fig:Cr3} are (almost) constant, the effect of the parameters $\chi_i$ can be notable for certain values of $\varphi_i$. 

Table \ref{Table:C} collects all $\Delta_aC_b$ and $\delta_aC_b$. Using this table, we compare the effect of two families of parameters, $\varphi_i$, $i=0,1,2$, and $\chi_i$, $i=1,2,3$. 
Note, that nonzero values $\Delta_{\chi_i}C_s$ appear because of leak of accuracy in calculating the integrals in (\ref{CC}). This table shows that all $\delta_aC_b$ are comparable between each other, while the variation  intervals $\Delta_{\varphi_i} C_b$ of mean values  $\langle C_b(\varphi_i)\rangle $  are much bigger than the variation
intervals   $\Delta_{\chi_i} C_b$. Consequently, the parameters $\varphi_i$ (describing the absolute values of probability amplitudes in the  pure sender's initial state) effect  mush stronger on the concurrence than the parameters $\chi_i$ (describing the phases of the above probability amplitudes). For this reason we refer to $\varphi_i$ and $\chi_i$  as strong and weak parameters respectively \cite{DZ_2017}.
 \begin{table}
\begin{tabular}{|c|cccc|}
\hline
$a$&$\Delta_a C_s$&$\delta_a C_s$& $\Delta_a C_r$&$\delta_a C_r$\cr
\hline
 $\varphi_0$&0.389 &$0.136\div 0.235$& 0.229&$0.081\div 0.208$ \cr
$\varphi_1$& 0.540 &$0\div 0.299$& 0.106&$0.057\div 0.226$ \cr
$\varphi_2$&0.499 &$0\div 0.296$& 0.386&$0.003\div 0.259$ \cr
\hline
 $\chi_1$& 0.001 &$0.227\div 0.229$&0&$0.186\div 0.187$ \cr
$\chi_2$& 0.001 &$0.227\div 0.230$&0&$0.186$ \cr
$\chi_3$& 0.001 &$0.227\div 0.230$& 0.018&$0.183\div 0.190$ \cr
\hline
\end{tabular}
\caption{List  of  $\Delta_a C_b$  and $\delta_a C_b$ for $b=s,r$,  $a=\varphi_i$, $i=0,1,2$ or $\chi_i$, $i=1,2,3$. The difference between strong parameters $\varphi_i$ and week parameters $\chi_i$ is shown: $\Delta_{\chi_i} C_b \ll \Delta_{\varphi_j} C_b$ for all pairs of $\chi_i$, $\varphi_j$, while $\delta_a C_b$ is almost the same for all cases.}
\label{Table:C}
\end{table}
 
\subsection{Restoring of only  $\pm2$-order coherence matrices}
\label{Section:example2}
As an example of a state including only even-order MQ-coherence matrices we consider the two-qubit matrix $s$ with 0  and $\pm 2$-order coherence matrices and restore only $2$-order coherence matrix (single element) ($(-2)$-order coherence matrix will be restored  automatically).
We use the same time instant for state registration $\tau_0=55.5352$ defined in Sec.\ref{Section:example1}. 
Since the excitation number is not conserved, then the element  $r_{14}$ is a linear combinations of the elements of 0, $\pm 2$-order MQ-coherence matrices  and therefore  it requires restoring. 

In this case we put  $\varphi_0=\varphi_1=\pi$, $\varphi_2=\varphi$, $\chi_3=\chi$ in (\ref{Favr}) co that only two elements in the initial state $|\psi\rangle$ are nonzero:
\begin{eqnarray}\label{Favr2}
 |\psi\rangle = \left(
 \begin{array}{c}
   \sin\frac{\varphi}{2}   \cr
0  \cr
0 \cr
 e^{i \chi} \cos\frac{\varphi}{2}
 \end{array}
 \right), \;\;s=|\psi\rangle \langle \psi| =\left(
\begin{array}{cccc}
s_{11}&0&0&s_{14}\cr
0&0&0&0\cr
0&0&0&0\cr
s_{14}^2&0&0&1-s_{11}\cr
\end{array}
\right),
 \end{eqnarray}
$s_{11}=\sin^2\frac{\varphi}{2}$, $s_{14}= e^{- i\chi} \sin\frac{\varphi}{2} \cos\frac{\varphi}{2}$.
 Now restoring system  (\ref{lameq}) includes 7 complex (or 14 real)  equations. 
Similar to Sec.\ref{Section:example1}, we find 1000 different solutions of the restoring system  and perform optimization over these solutions  according to condition  (\ref{opt2}).
 {\small \begin{eqnarray}\label{rhor12}
 &&r =\\\nonumber
 && \left(
 \begin{array}{cccc}
 r_{11} & 0 &0&  0.6681 e^{-0.6102 i}s_{14}\cr
0&r_{22}&r_{23}&0\cr
0&r^*_{23}&r_{33}&0\cr
0.6681 e^{0.6102 i}s^*_{14}&0&0&r_{44} 
 \end{array}
 \right)
 \end{eqnarray}}
 \begin{eqnarray}\nonumber
 r_{23}&=& 0.0256 e^{2.5773i}s_{11} + 0.2878 e^{1.8530i} s_{22} + 0.2743 e^{-1.7008i}s_{33} +\\\nonumber
 && 0.1398 e^{-3.1403 i}s_{44} +0.2536 e^{-1.0287 i}s_{23} + 0.2473 e^{1.1090 i}s^*_{23},\\\nonumber
 r_{11}&=& 0.8064s_{11} +   0.3888 s_{22} + 0.1923s_{33} + 0.0587s_{44} +2{\mbox{Re}} (0.1843 e^{-0.9604 i}s_{23}),\\\nonumber
  r_{22}&=&0.1384s_{11} +  0.2926  s_{22} +0.4564s_{33} +  0.1618s_{44} +2{\mbox{Re}} (0.2639  e^{0.9223 i}s_{23}),\\\nonumber
  r_{33}&=& 0.0468s_{11} +0.3056 s_{22} +  0.2643s_{33} + 0.1963s_{44} +2{\mbox{Re}} (0.2705  e^{-2.9240 i}s_{23}),\\\nonumber
 r_{44}&=&0.0084s_{11} +  0.0130 s_{22} + 0.0870s_{33} + 0.5832s_{44} \\\label{rhor02} 
&&+2{\mbox{Re}} (0.0013 e^{-2.3732 i}s_{23}).
  \end{eqnarray}
  In this case $N_r=2/16=1/8$, $\Lambda=\Lambda_{avr} = 0.6681$. 
Thus, the parameter $\Lambda$ is bigger than in Sec.\ref{Section:example1}. This is due to the fact that, for the fixed dimension of the extended receiver (that is for the fixed number $P$ of free parameters $\varphi_i$ in the unitary transformation), the variety of  solutions of restoring system (\ref{lameq}) increases with  decrease in the number of equations in it (this number is 70 in  Sec.\ref{Section:example1} and 14 in this section).

The fidelity of state $r$ (\ref{rhor12}) averaged over the sender's initial states (\ref{Favr2})  is $F=0.6569$, while the fidelity of the receiver's state before restoring is $F=0.7691$. This situation is similar to that in Sec.\ref{Section:ent1}: by restoring structure we reduce fidelity. To calculate fidelity in this section we use averaging over two parameters in the initial sender state (\ref{Favr2}):
\begin{eqnarray}
F =\int_0^{2 \pi} d\chi \int_0^\pi \langle \psi|r|\psi\rangle \sin \varphi\; d \varphi.
\end{eqnarray}
To avoid confusion we emphasize that only 2 elements of $r$ are restored. The $\pm 1$-order coherence matrices are  zero ($r^{(\pm 1)}=0$) not due to the restoring, but  because $s^{(\pm 1)}=0$ in the sender's initial state (\ref{Favr2}).

\subsubsection{Entanglement}
Similar, to Sec.\ref{Section:ent1}, as a quantity transferred from the sender to the receiver we consider the entanglements between two qubits of the sender and receiver in terms of concurrence (Wootters criterion) using 
 formulae (\ref{CR}) and (\ref{C}).
The concurrence $C_s$ can be easily calculated analytically: $C_s=\sin \varphi$, it does not depend on $\chi$. 
The concurrence $C_s$ is shown in Fig.\ref{Fig:Cr4}.

\begin{figure}[ht]
\centering
    \includegraphics[width=0.6 \textwidth]{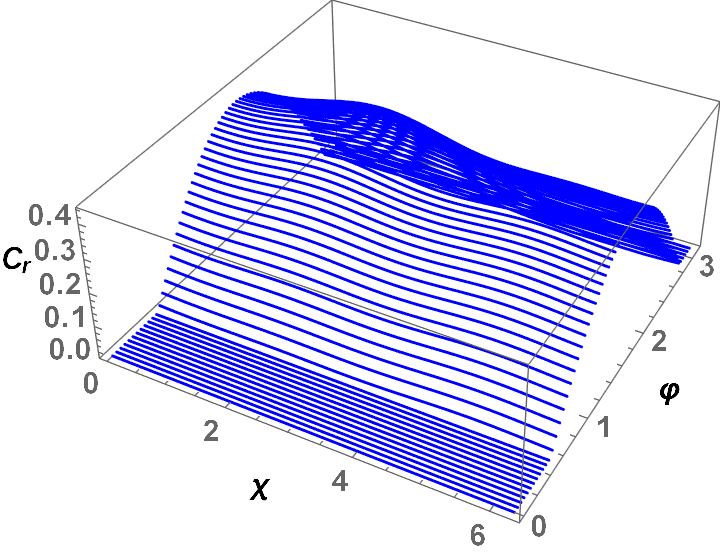}
    \caption{Concurrence $C_r$ over the plane  of parameters $\chi$ and $\varphi$.  }
    \label{Fig:Cr4}
  \end{figure}

In this case, there is no  region where $C_r>C_s$. 
In Fig.\ref{Fig:Cr4}, we see the region of zero concurrence, which is presented  on two-parametric plane $(\chi,\varphi)$ in  Fig. \ref{Fig:Cr5}. Notice that only two lines on $(\chi,\varphi)$-plane correspond to $C_s=0$: $\varphi =0$ and $\varphi=\pi$. 
Thus, the state-restoring decreases the concurrence. In particular, part of the states $s$ with  $C_s>0$ is mapped to the states $r$ with $C_r=0$. 

Fig.\ref{Fig:Cr4} demonstrates that the concurrence $C_r$ slightly depends on $\chi$, which is confirmed by Fig. \ref{Fig:Cr5}, where the region $C_r>0$ is separated from the regions $C_r=0$ by almost straight lines.

\begin{figure}[ht]
     \includegraphics[width=0.6\textwidth]{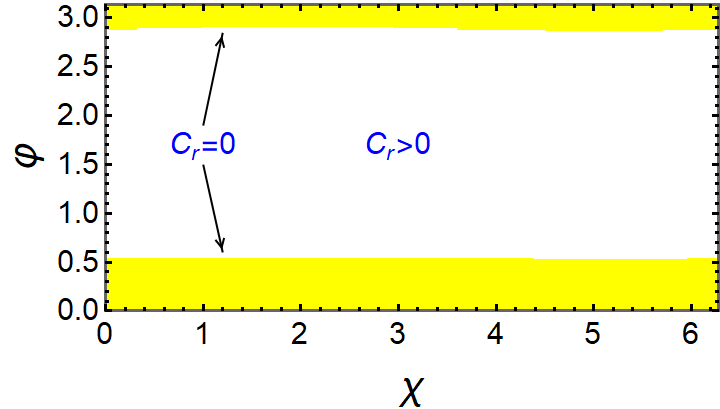}
     \caption{ The region  $C_r=0$  on the plane $(\chi,\varphi)$; $C_s=0$ only at $\varphi=0,\pi$.   }
    \label{Fig:Cr5}
  \end{figure}
 We don't consider the mean values and mean-square deviations of concurrences in this section because the two-parameter dependence of $C_s$ and $C_r$ on $\varphi$ and $\chi$ is well demonstrated, respectively,  in the analytical formula and  in Fig.\ref{Fig:Cr4} and Fig.\ref{Fig:Cr5}.

{\section{Discussion on  unitary restoring transformation}
\label{Section:discussion}

A principal object in the proposed algorithm is the unitary transformation of the  extended receiver, which has been first introduced in papers on remote state creation \cite{BZ_2016}. Its mathematical realization in terms of exponent of Hermitian matrices (\ref{exponent}) (see also \cite{Z_2018})  can be formally realized, for instance, using the Trotterization method \cite{Trotter,Suzuki}. However, a principal problem is the effective tool allowing to find the $\varphi$-parameters as solutions of the restoring system \cite{Z_2018,FPZ_2021}. Here, the optimization methods  (for instance, GRadient Ascent Pulse Engineering (GRAPE) method) with properly constructed target functions might be effective \cite{BFLPZ_2022,PP_2023,MP_2023}, although they usually allow to find a local extremum  instead of the global one. Nevertheless, multiple application of such method with subsequent choice of the best  result can yield acceptable output. 

A principal issue is losing the information in the process of state transfer because of   dispersion.   This is the reason of non-perfect restoring, i.e., all found $\lambda$-parameter are less then one by absolute value. Some increase in absolute values of $\lambda$-parameters can be reached by increasing the size of the extended receiver \cite{BFLPZ_2022} and, as a consequence,  the number of free parameters in the  unitary restoring transformation. 

A practical realization of the above control can be reached, in particular,  introducing the local inhomogeneous  magnetic field into the Hamiltonian \cite{FPZ_2024} with special step-wise dependence  on the time.  Then the parameters of unitary transformation are replaced by the Larmor frequencies in the Hamiltonian.  Although the problem of calculating the required Larmor frequencies remains in this case as well and it can be resolved via the mentioned above optimization methods.  Another natural tool of control is including the Lindblad terms with free parameters into the Liouville  evolution equation \cite{PP_2023,MP_2023} thus passing  to the master equation.  The mentioned free  parameters play the role of the parameters in the unitary restoring  transformation. Although these terms mean  interaction with environment and therefore usually lead to energy dissipation, nevertheless their combination with the above inhomogeneous magnetic  field could result in nontrivial manipulations with $\lambda$-parameters.

}

\section{Conclusions}
\label{Section:Conclusions}
We  represent the process of restoring the state transferred along  a spin chain governed by the XY-Hamiltonian that doesn't conserve the excitation number in the system. As a  consequence of this fact, we have to consider the  dynamics in the full  $2^n$-dimensional  space for the $n$-qubit system. We found that only MQ-coherence matrices of the same order-parity are mixed during evolution, which simplifies the restoring process. In addition, to prevent mixing of the multiple-quantum coherence matrices of different order-parity after applying the  unitary restoring transformation, this transformation must include only the blocks  which change the excitation number in the  density matrix by  an even value. 

As a time instant for state registration, we select such instant that maximizes the fidelity of a pure state transfer (before applying the  unitary restoring transformation) averaged over all pure initial sender's states. 

We notice that the fidelity of the restored state transfer doesn't characterize the effectivity of restoring protocol. Therefore we introduce three characteristics describing the features of the restored state. These are the part of restored elements $N_r$, the minimal by absolute value $\lambda$-parameter $\Lambda$ and the mean absolute value of  $\lambda$-parameters  $\Lambda_{avr}$.

We consider an example of two-qubit state restoring  in the six-qubit chain.
For that aim we involve the four-qubit extended receiver and apply the  unitary restoring transformation to it.
Optimization of restoring was performed over 1000 different solutions of restoring system (\ref{lameq}).
Studying  the concurrence as a measure of entanglement transferred from the  sender to the receiver we evaluate   100000 random initial states of the sender  and show that the concurrence can be amplified (i.e., $C_r>C_s$) provided that the sender's concurrence doesn't exceed 0.561.  
 
The evolution of  a two-qubit  state including only the even-order MQ-coherence matrices  is considered as a simplest example. In this  case, the pure initial sender's state depends on two parameters which simplifies the visualization of the results, in particular, understanding the concurrence between qubits of the sender and receiver. 

Studying the  above examples of the two-qubit restoring, we conclude that the parameters of the pure sender's initial state have different influence on the concurrence in the sender and receiver. The parameters responsible for the absolute values of the probability amplitudes  in the sender's state (parameters $\varphi_i$) effect on the concurrence 
much stronger than the parameters responsible for the phases of those probability amplitudes (parameters $\chi_i$). We refer to these parameters as strong and weak parameters respectively, 

{\bf Acknowledgments.} The work was performed as a part of a state task, State Registration
No. 124013000760-0.

\section{Appendix: $W$-representation of $\lambda$-parameters}
\label{Section:Appendix}
To represent the $\lambda$-parameters in  (\ref{rr}) in terms of the elements of the unitary transformation $W$ given in (\ref{W}), we introduce multi-indexes \cite{SZ_2019}. Thus, capital Latin letters with subscripts,  $I_S$, $I_{TL}$, $I_R$, 
are used for labeling the nodes of, respectively, the sender, transmission line and receiver. We assume the binary representation for these multi-indexes.
Then Eq.(\ref{rr}),  which follows from Eq. (\ref{rhot3}), can be given the following form:
\begin{eqnarray}
&&
r^{(nm)}_{N_R;M_R} =\\\nonumber
&& \sum_{{N_S,N_{TL}}\atop{I_S,J_S}}  \sum_{ l- k = p \!\!\!\!\! \mod \!2}  W^{(nk)}_{N_SN_{TL}N_R; I_S0_{TL}0_R} s^{(kl)}_{I_S;J_S} (W^\dagger)^{(lm)}_{J_S0_{TL} 0_R;N_SN_{TL}M_R},
 \end{eqnarray}
where $n,\;m,\;k,\;l$ satisfy (\ref{nmjk}). Here the first sum is over such indexes $N_S$, $N_{TL}$, 
$I_S$, $J_S$ that  
\begin{eqnarray}\nonumber
&&
|I_S|-(|N_S|+|N_{TL}|+|N_R|) = n-k =0\!\!\!\mod 2,\\\nonumber
&&
 |J_S|-(|N_S|+|N_{TL}|+|M_R| )= m-l =0\!\!\!\mod 2,\\\label{MI}
 &&
 |J_S|-|I_S| = l-k \!\!\!\mod 2,
\end{eqnarray}
where $|J|$ means the number of units in the multi-index $J$.
Thus,  for $\lambda$-parameters in (\ref{rr}) we have
\begin{eqnarray}
\lambda_{N_RM_RI_SJ_S}^{(nmkl)}= \sum_{N_S,N_{TL}}   W^{(nk)}_{N_SN_{TL}N_R; I_S0_{TL}0_R}  (W^\dagger)^{(lm)}_{J_S0_{TL} 0_R;N_SN_{TL}M_R},
\end{eqnarray}
with relations (\ref{nmjk}), $l-k=p\!\!\!\mod 2$ and  (\ref{MI})  among superscripts and multi-indexes respectively.

\end{document}